\documentclass[11pt]{article}

\usepackage{hyperref,amsfonts,amsmath,amssymb,graphicx,bm,subfigure,a4wide}

\numberwithin{equation}{section}

\begin{document}

\title{\textbf{Interaction of inhomogeneous axions with magnetic fields in the early universe}}

\author{Maxim Dvornikov\thanks{maxdvo@izmiran.ru}
\\
\small{\ Pushkov Institute of Terrestrial Magnetism, Ionosphere} \\
\small{and Radiowave Propagation (IZMIRAN),} \\
\small{108840 Moscow, Troitsk, Russia}}

\date{}

\maketitle

\begin{abstract}
We study the system of interacting axions and magnetic fields in the
early universe after the quantum chromodynamics phase transition,
when axions acquire masses. Both axions and magnetic fields are supposed
to be spatially inhomogeneous. We derive the equations for the spatial
spectra of these fields, which depend on conformal time. In case of
the magnetic field, we deal with the spectra of the energy density
and the magnetic helicity density. The evolution equations are obtained
in the closed form within the mean field approximation. We choose
the parameters of the system and the initial condition which correspond
to realistic primordial magnetic fields and axions. The system of
equations for the spectra is solved numerically. We compare the cases of  inhomogeneous and homogeneous axions. The evolution of the magnetic field in these cases is different only within small time intervals. Generally, magnetic fields are driven mainly by the magnetic diffusion. We find that the magnetic field instability takes place for the amplified initial wavefunction of the homogeneous axion. This instability is suppressed if we account for the  inhomogeneity of the axion.
\end{abstract}

\section{Introduction}

The existence of axions, which are pseudoscalar particles, was theoretically
proposed in Ref.~\cite{PecQui77} to provide the conservation of
the CP symmetry in the quantum chromodynamics (QCD). The conserved
CP symmetry is required, in particular, because of the experimental
non-observation of an electric dipole moment of a neutron (see, e.g.,
Ref.~\cite{Abe20}). There are numerous studies of the models for
QCD axions which are summarized in Ref.~\cite{Luz20}.

Currently, there are active searches for axions in laboratory experiments.
These particles can be revealed in their interactions with electromagnetic
fields, fermion spins, and nuclear electric dipole moments. The difficulty
of a direct axion observation is because of the smallness of corresponding
coupling constants. Major present experimental studies of axions are
reviewed in Ref.~\cite{Gra15}. There are recent claims (see, e.g.,
Ref.~\cite{Apr20}) that some experimental data can be interpreted
as the axions observation. However, a more careful analysis to confirm
such findings is required.

Besides laboratory studies, there are attempts to detect axions from
astrophysical sources and cosmological media. One can try to capture
axions produced in the Sun using existing and future helioscopes (see,
e.g., Ref.~\cite{Ana17}). Additionally, axions created in neutron
stars can result in their X-ray emission~\cite{Bus21}, which is
observed by satellite based telescopes. Axions, which the cold dark
matter halo in our Galaxy consists of, can be detected using a haloscope
device. There are many haloscopes techniques, which are described
in Ref.~\cite{SemYou21}.

The interest to the experimental study of astrophysical and cosmological
axions is motivated by the fact that these particles can both solve
the CP symmetry problem in QCD and be one of the most plausible candidate
for the dark matter~\cite{Cha21}. For example, an axion can be the candidate for the dark matter using the misalignment idea~\cite{DinFis83}. There are numerous sources of
axions in the early universe. These particles can be produced in decays
of more massive particles or topological defects, be thermally emitted
by other constituents of the standard model etc. Some of these mechanisms
are reviewed in Ref.~\cite{Mar16}.

Cosmological axions are believed to be spatially homogeneous. However,
if the Peccei-Quinn phase transition, which takes place at $T_{\mathrm{PQ}}\approx f_{a}=(10^{10}-10^{12})\,\text{GeV}$,
where $f_{a}$ is the Peccei-Quinn constant, happens after the reheating
during the inflation, then nonzero spatial modes of axions can appear.
Such a scenario is also not excluded (see, e.g., Refs.~\cite{Cha21,Mar16}).
Inhomogeneous axions were proposed in Ref.~\cite{KolTka93} to explain the formation of Bose stars. The contribution of inhomogeneous axions to the cold dark matter was discussed in Ref.~\cite{KhlSakSok99}. The evolution of axion miniclusters was recently studied in Ref.~\cite{EnaParSch17}.

In the present work, we study the evolution of the system of cosmological
axions and magnetic fields. For the first time, analogous problem
was discussed in Ref.~\cite{LonVac15}. However, nontrivial spectra
of neither axions nor magnetic fields were considered in Ref.~\cite{LonVac15}.
Moreover, the universe expansion was not accounted for properly there.
In Ref.~\cite{DvoSem20}, we refined the analysis of Ref.~\cite{LonVac15}
by considering the spectral distribution of a primordial magnetic
field and taking into account the time dependent scale factor $a\neq1$.
Now, we generalize the results of Ref.~\cite{DvoSem20} by accounting
for an arbitrary spatial spectrum of axions.

As mentioned in Refs.~\cite{LonVac15,DvoSem20}, a magnetic field
in the presence of an axion can be unstable. Indeed, in this case,
the induction equation gets the contribution, $\propto(\nabla\times\mathbf{B})$,
analogous to that for the chiral magnetic effect (the CME)~\cite{FukKhaWar08}.
The axion wavefunction contributes to the $\alpha$-dynamo parameter
responsible for the magnetic field instability. Thus, the consideration
of inhomogeneous axions is equivalent to the study of chiral media
with a coordinate dependent chiral imbalance. This problem was studied
in Ref.~\cite{Bra17}. However, unlike the CME case with a coordinate
dependent chiral imbalance, here, we have a well established wave
equation for the axion wavefunction interacting with an electromagnetic
field.

Our work is organized in the following way. In Sec.~\ref{sec:MAGNAX},
we derive the general equations for the spectra of the magnetic energy
and helicity densities in the presence of an inhomogeneous axion.
The equation for the Fourier transform of the axion wavefunction was
considered in Sec.~\ref{sec:AXMAGN}. Applying the mean field approximation,
we combine the obtained equations to the closed system in Sec.~\ref{sec:MEAN}.
In Sec.~\ref{sec:INICOND}, we choose the parameters of the system, corresponding
to the realistic cosmological medium, and define the initial condition.
We present the results of the numerical simulations in Sec.~\ref{sec:RES}. In Sec.~\ref{subsec:TOY}, we study the possibility to trigger the magnetic field instability in the system.
Finally, we conclude in Sec.~\ref{sec:CONCL}.

\section{Magnetic field in the presence of inhomogeneous axions\label{sec:MAGNAX}}

The equations for the electromagnetic field $F_{\mu\nu}$, interacting
with a scalar axion field $\varphi$, in a curved spacetime with the
coordinates $x^{\mu}=(t,\mathbf{x})$ read
\begin{align}\label{eq:Maxcov}
  \frac{1}{\sqrt{-g}}\partial_{\nu}(\sqrt{-g}F^{\mu\nu})+g_{a\gamma}\partial_{\nu}\varphi\tilde{F}^{\mu\nu}+J^{\mu} & =0,
  \nonumber
  \\
  \frac{1}{\sqrt{-g}}\partial_{\nu}(\sqrt{-g}\tilde{F}^{\mu\nu}) & =0,
\end{align}
where $\tilde{F}^{\mu\nu}=\tfrac{1}{2}E^{\mu\nu\alpha\beta}F_{\alpha\beta}$
is the dual counterpart of $F_{\mu\nu}$, $E^{\mu\nu\alpha\beta}=\tfrac{1}{\sqrt{-g}}\varepsilon^{\mu\nu\alpha\beta}$
is the covariant antisymmetric tensor, $\varepsilon^{0123}=+1$, $g=\det(g_{\mu\nu})$,
$g_{\mu\nu}$ is the metric tensor, $g_{a\gamma}$ is the coupling
constant, and $J^{\mu}=(\rho,\mathbf{J}/a)$ is the external current.
We consider the Friedmann-Robertson-Walker (FRW) metric $g_{\mu\nu}=\text{diag}(1,-a^{2},-a^{2},-a^{2}),$where
$a(t)$ is the scale factor. Using the conformal variables~\cite{BraEnqOle96}
$\mathbf{E}_{c}=a^{2}\mathbf{E}$, $\mathbf{B}_{c}=a^{2}\mathbf{B}$,
$\rho_{c}=a^{3}\rho$, and $\mathbf{J}_{c}=a^{3}\mathbf{J}$, we rewrite
Eq.~(\ref{eq:Maxcov}) in the form,
\begin{align}
  (\nabla\times\mathbf{B}_{c}) & =
  \mathbf{E}_{c}'+\mathbf{J}_{c}+g_{a\gamma}\varphi'\mathbf{B}_{c}+g_{a\gamma}(\nabla\varphi\times\mathbf{E}_{c}),
  \label{eq:Max1}
  \\
  (\nabla\times\mathbf{E}_{c}) & =-\mathbf{B}_{c}',
  \label{eq:Max2}
  \\
  (\nabla\mathbf{E}_{c}) & =-g_{a\gamma}(\mathbf{B}_{c}\nabla)\varphi+\rho_{c},
  \label{eq:Max3}
  \\
  (\nabla\mathbf{B}_{c}) & =0,
\end{align}
where the prime means the derivative with respect to the conformal
time $\eta$ defined by $\mathrm{d}t=a\mathrm{d}\eta$.

Supposing that $\mathbf{J}_{c}=\sigma_{c}\mathbf{E}_{c}$, where $\sigma_{c}=10^{2}$
is the conformal conductivity of ultrarelativistic plasma, and neglecting
the displacement current $\mathbf{E}_{c}'$ in Eq.~(\ref{eq:Max1}),
we get that the electric field has the form,
\begin{equation}\label{eq:Elfield}
  \mathbf{E}_{c}=\frac{1}{\sigma_{c}}
  \left[
    (\nabla\times\mathbf{B}_{c})-g_{a\gamma}\varphi'\mathbf{B}_{c}
  \right]-
  \frac{1}{\sigma_{c}^{2}}
  \left[
    g_{a\gamma}
    \left(
      \nabla\varphi\times(\nabla\times\mathbf{B}_{c})
    \right)+
    g_{a\gamma}^{2}\varphi'(\mathbf{B}_{c}\times\nabla\varphi)
  \right].
\end{equation}
Here we keep the terms linear in $\nabla\varphi$. Setting $\rho_{c}=0$
in Eq.~(\ref{eq:Max3}), we derive the equation for $\mathbf{B}_{c}$,
\begin{align}\label{eq:Indgen}
  \mathbf{B}_{c}'= & \frac{1}{\sigma_{c}}
  \left[
    \Delta\mathbf{B}_{c}+g_{a\gamma}\varphi'(\nabla\times\mathbf{B}_{c})+g_{a\gamma}(\nabla\varphi'\times\mathbf{B}_{c})
  \right]
  \nonumber
  \\
  & +
  \frac{g_{a\gamma}}{\sigma_{c}^{2}}
  \left\{
    \left(
      (\nabla\times\mathbf{B}_{c})\cdot\nabla
    \right)
    \nabla\varphi-
    \left(
      \nabla\varphi\cdot\nabla
    \right)
    (\nabla\times\mathbf{B}_{c})-(\nabla\times\mathbf{B}_{c})\Delta\varphi
  \right\}
  \nonumber
  \\
  & +
  \frac{g_{a\gamma}^{2}}{\sigma_{c}^{2}}
  \big\{
    \mathbf{B}_{c}
    \left[
      \varphi'\Delta\varphi+(\nabla\varphi'\cdot\nabla\varphi)
    \right]
    \notag
    \\
    & +
    \varphi'
    \left[
      (\nabla\varphi\cdot\nabla)\mathbf{B}_{c}-(\mathbf{B}_{c}\cdot\nabla)\nabla\varphi
    \right]-
    \nabla\varphi(\mathbf{B}_{c}\cdot\nabla)\varphi'
  \big\},
\end{align}
where we use Eqs.~(\ref{eq:Max2}) and~(\ref{eq:Elfield}).

Equation~(\ref{eq:Indgen}) can be simplified if we assume that axions
are inhomogeneous but isotropic. It means that, after the spatial
averaging, one has that $\left\langle \nabla\varphi\right\rangle =0$,
whereas $\left\langle (\nabla\varphi)^{2}\right\rangle \neq0$ and
$\left\langle \Delta\varphi\right\rangle \neq0$. Keeping the terms
linear in $g_{a\gamma}$ in Eq.~(\ref{eq:Indgen}) and spatially
averaging it, we get the modified induction equation,
\begin{align}\label{eq:Ind}
  \mathbf{B}_{c}'= & \frac{1}{\sigma_{c}}\Delta\mathbf{B}_{c}+\alpha(\mathbf{x})(\nabla\times\mathbf{B}_{c}),
  \quad
  \alpha(\mathbf{x})=\frac{g_{a\gamma}}{\sigma_{c}}
  \left(
    \varphi'-\frac{2}{3\sigma_{c}}\Delta\varphi
  \right),
\end{align}
where we omit the averaging brackets for the sake of brevity. In Eq.~(\ref{eq:Ind}),
the $\alpha$-dynamo parameter $\alpha(\mathbf{x})$, which is the
coefficient in the magnetic instability term $\propto(\nabla\times\mathbf{B}_{c})$,
is different from that in Refs~\cite{LonVac15,DvoSem20}. It gets
the contribution from the inhomogeneous axion field $\propto\Delta\varphi$.

It is convenient to deal with a Fourier transform, which has the form,
$f_{\mathbf{k}}=\smallint\mathrm{d}^{3}xe^{\mathrm{i}\mathbf{kx}}f(\mathbf{x}),$
and $f_{\mathbf{k}}^{*}=f_{-\mathbf{k}}$ since $f(\mathbf{x})$ is
real. Then we define the spectra of the magnetic energy density and
the magnetic helicity,
\begin{align}\label{eq:specgen}
  \rho_{\mathbf{p}}(\eta) & =\frac{1}{2}\int\mathrm{d}^{3}xe^{\mathrm{i}\mathbf{px}}\mathbf{B}_{c}^{2}(\mathbf{x})=
  \frac{1}{2}\int\frac{\mathrm{d}^{3}k}{(2\pi)^{3}}(\mathbf{B}_{\mathbf{k}}^{(c)*}\mathbf{B}_{\mathbf{k}+\mathbf{p}}^{(c)}),
  \nonumber
  \\
  h_{\mathbf{p}}(\eta) & =\int\mathrm{d}^{3}xe^{\mathrm{i}\mathbf{px}}(\mathbf{A}_{c}(\mathbf{x})\mathbf{B}_{c}(\mathbf{x}))=
  \int\frac{\mathrm{d}^{3}k}{(2\pi)^{3}}(\mathbf{A}_{\mathbf{k}}^{(c)*}\mathbf{B}_{\mathbf{k}+\mathbf{p}}^{(c)}).
\end{align}
Note that the total magnetic helicity,
\begin{equation}\label{eq:heldef}
  H(\eta)=\int\mathrm{d}^{3}x(\mathbf{A}_{c}\mathbf{B}_{c})=
  \int\frac{\mathrm{d}^{3}p}{(2\pi)^{3}}(\mathbf{A}_{\mathbf{p}}^{(c)}\mathbf{B}_{\mathbf{p}}^{(c)*}),
\end{equation}
has the standard form.

Taking the Fourier transform of Eq.~(\ref{eq:Ind}) and forming the
binary combinations of the fields, one gets the following differential
equations:
\begin{align}\label{eq:speceqgen}
  \rho'_{\mathbf{q}}= & -\frac{1}{2\sigma_{c}}
  \int\frac{\mathrm{d}^{3}k}{(2\pi)^{3}}k^{2}
  \left[
    (\mathbf{B}_{\mathbf{k}}^{(c)*}\mathbf{B}_{\mathbf{k}+\mathbf{q}}^{(c)})+
    (\mathbf{B}_{\mathbf{k}}^{(c)}\mathbf{B}_{\mathbf{k}-\mathbf{q}}^{(c)*})
  \right]
  \nonumber
  \\
  & +
  \frac{1}{2}\int\frac{\mathrm{d}^{3}p}{(2\pi)^{3}}\frac{\mathrm{d}^{3}k}{(2\pi)^{3}}p^{2}
  \left[
    \alpha_{\mathbf{p}-\mathbf{k}}^{*}\mathbf{A}_{\mathbf{p}}^{(c)}\mathbf{B}_{\mathbf{k}-
    \mathbf{q}}^{(c)*}+\alpha_{\mathbf{p}-\mathbf{k}}\mathbf{A}_{\mathbf{p}}^{(c)*}\mathbf{B}_{\mathbf{k}+\mathbf{q}}^{(c)}
  \right],
  \nonumber
  \\
  h'_{\mathbf{q}}= & -\frac{1}{\sigma_{c}}
  \int\frac{\mathrm{d}^{3}k}{(2\pi)^{3}}k^{2}
  \left[
    (\mathbf{A}_{\mathbf{k}}^{(c)*}\mathbf{B}_{\mathbf{k}+\mathbf{q}}^{(c)})+
    (\mathbf{A}_{\mathbf{k}-\mathbf{q}}^{(c)*}\mathbf{B}_{\mathbf{k}}^{(c)})
  \right]
  \nonumber
  \\
  & +
  \int\frac{\mathrm{d}^{3}p}{(2\pi)^{3}}\frac{\mathrm{d}^{3}k}{(2\pi)^{3}}
  \bigg\{
    \alpha_{\mathbf{p}-\mathbf{k}}^{*}\mathrm{i}
    \left(
      \mathbf{p}\times\mathbf{A}_{\mathbf{k}-\mathbf{q}}^{(c)*}
    \right)
    \mathbf{B}_{\mathbf{p}}^{(c)}
    \notag
    \\
    & +
    \alpha_{\mathbf{p}-\mathbf{k}}\frac{1}{k^{2}}
    \left[
      (\mathbf{kp})
      \left(
        \mathbf{B}_{\mathbf{p}}^{(c)*}\mathbf{B}_{\mathbf{k}+\mathbf{q}}^{(c)}
      \right)-
      \left(
        \mathbf{p}\mathbf{B}_{\mathbf{k}+\mathbf{q}}^{(c)}
      \right)
      \left(
        \mathbf{k}\mathbf{B}_{\mathbf{p}}^{(c)*}
      \right)
    \right]
  \bigg\},
\end{align}
for the spectra $\rho_{\mathbf{p}}(\eta)$ and $h_{\mathbf{p}}(\eta)$
in Eq.~(\ref{eq:specgen}). In Eq.~(\ref{eq:speceqgen}), $\alpha_{\mathbf{k}}=\smallint\mathrm{d}^{3}xe^{\mathrm{i}\mathbf{kx}}\alpha(\mathbf{x})$
is the Fourier transform of the $\alpha$-dynamo parameter in Eq.~(\ref{eq:Ind}).
We mention that the equations for the spectra in Eq.~(\ref{eq:specgen})
cannot be written in a closed form, analogous to that, e.g., in Refs.~\cite{BoyFroRuc12,DvoSem14},
for a general coordinate dependent $\alpha$-dynamo parameter.

\section{Axions in the presence of an electromagnetic field\label{sec:AXMAGN}}

The evolution of an axion field under the influence of an electromagnetic
field obeys the following equation in a curved spacetime:
\begin{equation}\label{eq:axcurved}
  \frac{1}{\sqrt{-g}}\partial_{\mu}\left(\sqrt{-g}\partial^{\mu}\varphi\right)+m^{2}\varphi+
  \frac{g_{a\gamma}}{4}F_{\mu\nu}\tilde{F}^{\mu\nu}=0,
\end{equation}
where $m$ is the axion mass. We can rewrite Eq.~(\ref{eq:axcurved}) in the
form,
\begin{equation}\label{eq:axconf}
  \varphi^{\prime\prime}+\frac{2a'}{a}\varphi'-\nabla^{2}\varphi+a^{2}m^{2}\varphi=
  \frac{g_{a\gamma}}{a^{2}}(\mathbf{E}_{c}\mathbf{B}_{c}),
\end{equation}
where we take the FRW metric and use the conformal variables.

In the Fourier representation, Eq.~(\ref{eq:axconf}) takes the form, 
\begin{equation}\label{eq:axfour}
  \varphi_{\mathbf{k}}^{\prime\prime}+\frac{2a'}{a}\varphi_{\mathbf{k}}'+k^{2}\varphi_{\mathbf{k}}+a^{2}m^{2}\varphi_{\mathbf{k}}=
  \frac{g_{a\gamma}}{a^{2}}
  \int\frac{\mathrm{d}^{3}p}{(2\pi)^{3}}(\mathbf{E}_{\mathbf{p}+\mathbf{k}}^{(c)}\mathbf{B}_{\mathbf{p}}^{(c)*}).
\end{equation}
The integral in the right hand side of Eq.~(\ref{eq:axfour}) differs
from the derivative of the magnetic helicity,
\begin{equation}\label{eq:helsimint}
  \int\frac{\mathrm{d}^{3}p}{(2\pi)^{3}}(\mathbf{E}_{\mathbf{p}+\mathbf{k}}^{(c)}\mathbf{B}_{\mathbf{p}}^{(c)*})\neq
  -\frac{1}{2}H'(\eta)=\int\mathrm{d}^{3}x(\mathbf{E}{}_{c}\mathbf{B}_{c})=
  \int\frac{\mathrm{d}^{3}p}{(2\pi)^{3}}(\mathbf{E}_{\mathbf{p}}^{(c)}\mathbf{B}_{\mathbf{p}}^{(c)*}),
\end{equation}
which is given in Eq.~(\ref{eq:heldef}).

Nevertheless, we assume that the electromagnetic field is inhomogeneous
but isotropic. Thus all odd correlators, like $\left\langle k_{i}\right\rangle $,
$\left\langle k_{i}k_{j}k_{n}\right\rangle $ etc., vanish. In this
case, we rewrite Eq.~(\ref{eq:helsimint}) as
\begin{equation}\label{eq:helfkp}
  \int\frac{\mathrm{d}^{3}p}{(2\pi)^{3}}(\mathbf{E}_{\mathbf{p}+\mathbf{k}}^{(c)}\mathbf{B}_{\mathbf{p}}^{(c)*})=
  \int\mathrm{d}^{3}x\exp\left(-\frac{k^{2}x^{2}}{6}\right)(\mathbf{E}_{c}(\mathbf{x})\mathbf{B}_{c}(\mathbf{x}))=
  -\frac{1}{2}\int\frac{\mathrm{d}^{3}p}{(2\pi)^{3}}h'_{\mathbf{p}}(\eta)f_{\mathbf{p}}^{(k)*},
\end{equation}
where $h_{\mathbf{p}}(\eta)$ is the spectrum of the magnetic helicity
density in Eq.~(\ref{eq:specgen}) and
\begin{equation}\label{eq:fkp}
  f_{\mathbf{p}}^{(k)}=\int\mathrm{d}^{3}xe^{\mathrm{i}\mathbf{px}}\exp
  \left(
    -\frac{k^{2}x^{2}}{6}
  \right)=
  \frac{6\sqrt{6}\pi^{3/2}}{k^{3}}\exp
  \left(
    -\frac{3p^{2}}{2k^{2}}
  \right),
\end{equation}
is the scalar function.

\section{Mean field approximation\label{sec:MEAN}}

Equations~(\ref{eq:speceqgen}) and~(\ref{eq:axfour}) are difficult
to analyze in general case. Thus, we apply the mean field approximation.
It implies the consideration of nontrivial spectra which are attributed,
however, to the zero mode only.

First, we assume in Eq.~(\ref{eq:speceqgen}) that $\alpha_{\mathbf{k}}=\alpha_{0}\delta^{3}(\mathbf{k})$,
where
\begin{equation}\label{eq:alpha0}
  \alpha_{0}(\eta)=\int\mathrm{d}^{3}k\alpha_{\mathbf{k}}(\eta)=\frac{4\pi g_{a\gamma}}{\sigma_{c}}\int k{}^{2}\mathrm{d}k
  \left[
    \varphi'_{k}(\eta)+\frac{2k^{2}}{3\sigma_{c}}\varphi_{k}(\eta)
  \right].
\end{equation}
In this approximation, we set $\mathbf{q}=0$ in Eq.~(\ref{eq:speceqgen})
and consider the mean spectra of the magnetic energy and the magnetic
helicity, $\rho(k,\eta)=\tfrac{k^{2}}{4\pi^{2}}(\mathbf{B}_{\mathbf{k}}^{(c)*}\mathbf{B}_{\mathbf{k}}^{(c)})$
and $h(k,\eta)=\tfrac{k^{2}}{2\pi^{2}}(\mathbf{A}_{\mathbf{k}}^{(c)*}\mathbf{B}_{\mathbf{k}}^{(c)})$,
which were introduced, e.g., in Refs.~\cite{BoyFroRuc12,DvoSem14}.
They are related to the total spectra by the expressions,
\begin{equation}\label{eq:rho0h0}
  \rho_{\mathbf{q}=0}=\int\mathrm{d}k\rho(k,\eta),
  \quad
  h_{\mathbf{q}=0}=\int\mathrm{d}kh(k,\eta),
\end{equation}
where $\rho_{\mathbf{p}}$ and $h_{\mathbf{p}}$ are given in Eq.~(\ref{eq:specgen}).
Using Eq.~(\ref{eq:speceqgen}) in the considered approximation,
we get that the mean spectra obey the equations,
\begin{align}\label{eq:speceqmean}
  \frac{\partial\rho(k,\eta)}{\partial\eta} & =-\frac{2k^{2}}{\sigma_{c}}\rho(k,\eta)+\alpha_{0}k^{2}h(k,\eta),
  \nonumber
  \\
  \frac{\partial h(k,\eta)}{\partial\eta} & =-\frac{2k^{2}}{\sigma_{c}}h(k,\eta)+4\alpha_{0}\rho(k,\eta),
\end{align}
which formally coincide with those derived in Refs.~\cite{BoyFroRuc12,DvoSem14}.
However, Eq.~(\ref{eq:speceqmean}) involves a nontrivial spectrum
of the axion field in Eq.~(\ref{eq:alpha0}).

Then, we consider the limit $k\to0$ in the integrand in Eq.~(\ref{eq:axfour}).
In this case, $f_{\mathbf{p}}^{(k\to0)}=(2\pi)^{3}\delta^{3}(\mathbf{p})$
in Eq.~(\ref{eq:fkp}). Using Eq.~(\ref{eq:helfkp}) in this approximation
and Eq.~(\ref{eq:rho0h0}), we rewrite Eq.~(\ref{eq:axfour}) in
the form,
\begin{equation}\label{eq:axeqmean}
  \varphi_{k}^{\prime\prime}+\frac{2a'}{a}\varphi_{k}'+k^{2}\varphi_{k}+a^{2}m^{2}\varphi_{k}=
  -\frac{g_{a\gamma}}{2a^{2}}\int\mathrm{d}k'\frac{\partial h(k',\eta)}{\partial\eta},
\end{equation}
which is similar to that used in Ref.~\cite{DvoSem20}. However,
we keep the term $\propto k^{2}\varphi_{k}$ in the left hand side of Eq.~\eqref{eq:axeqmean}.

Equations~(\ref{eq:axeqmean}) and~(\ref{eq:speceqmean}), together
with Eq.~(\ref{eq:alpha0}), form the closed set of the evolution
equations, which define the evolution of the magnetic and axion fields.
We shall solve them numerically in Secs.~\ref{sec:INICOND} and~\ref{sec:RES}.

\section{Initial condition and the parameters of the system\label{sec:INICOND}}

We consider the evolution of the system after the QCD phase transition
which happens at $T_{\mathrm{QCD}}=160\,\text{MeV}$. In this situation,
the axion mass is taken to be constant, $m=10^{-6}\,\text{eV}$. We
use the parameterization in which the scale factor $a=1/T$ is dimensional,
where $T$ is the primordial plasma temperature. The conformal time is defined
up to a constant. We take that $\eta=\tilde{M}_{\mathrm{Pl}}(T^{-1}-T_{\mathrm{QCD}}^{-1})$,
where $\tilde{M}_{\mathrm{Pl}}=M_{\mathrm{Pl}}/1.66\sqrt{g_{*}}$,
$M_{\mathrm{Pl}}=1.2\times10^{19}\,\text{GeV}$ is the Planck mass,
and $g_{*}=17.25$ is the number of the relativistic degrees of freedom
at $T_{\mathrm{QCD}}$~\cite{Hus16}. In this case $\eta(T_{\mathrm{QCD}})=0$.

We assume that the seed spectrum of the magnetic energy is Kolmogorov,
$\rho_{0}(k)\propto k^{n}$ with $n=-5/3$. The seed spectrum of the
magnetic helicity is $h_{0}(k)=2q\rho_{0}(k)/k$, where $0\leq q\leq1$
is the parameter fixing the initial helicity. Note that $k$ is the
dimensionless conformal momentum. It spans the range $k_{\mathrm{min}}<k<k_{\mathrm{max}}$,
where $k_{\mathrm{min}}=T_{\mathrm{QCD}}/\tilde{M}_{\mathrm{Pl}}=9.1\times10^{-20}$
is the reciprocal horizon size at $T_{\mathrm{QCD}}$. The maximal
conformal momentum $k_{\mathrm{max}}$ is a free parameter in our
model. We take that $k_{\mathrm{max}}<r_{\mathrm{D}}^{-1}=0.1$, where
$r_{\mathrm{D}}$ is the conformal Debye length, to guarantee the
plasma electroneutrality.

There is not so much knowledge about the seed axion spectrum. We assume
that $\varphi_{k}^{(0)}=8\pi^{3/2}f_{a}\exp(-k^{2}/\delta^{2})/\delta^{3}$,
where $\delta$ is the phenomenological parameter. There is the approximate
relation between $m$ and $f_{a}$~\cite{Cha21}: $m=(5.7\pm0.007)\,\mu\text{eV}\left(\tfrac{10^{12}\,\text{GeV}}{f_{a}}\right)$.
Thus, we take $f_{a}=10^{12}\,\text{GeV}$ since we have chosen $m=10^{-6}\,\text{eV}$.
The suggested values of $m$ and $f_{a}$ are in the sensitivity range
of the operating and future lumped element detectors~\cite{SemYou21}
(see also Refs.~\cite{Oue19,Gra21}). We take that $0\leq k<k_{\mathrm{max}}$
in the axion seed spectrum to be able to probe small momenta. Note
that $\varphi_{k}^{(0)}\to(2\pi)^{3}f_{a}\delta^{3}(\mathbf{k})$
at $\delta\to0$. In this case, we can reproduce the zero mode approximation
studied in Ref.~\cite{DvoSem20}. We shall consider $\delta\sim k_{\mathrm{max}}$
as a rule. Following Ref.~\cite{DvoSem20}, we estimate the initial
first derivative $\partial_{\eta}\varphi_{k}^{(0)}$ basing on the
virial theorem. Computing the zero component of the energy-momentum
tensor of an axion, we get that $\partial_{\eta}\varphi_{k}^{(0)}=\sqrt{k^{2}+m^{2}/T_{\mathrm{QCD}}^{2}}\varphi_{k}^{(0)}$.

We introduce the new variables in Eqs.~(\ref{eq:speceqmean}) and~(\ref{eq:axeqmean}),
\begin{align}\label{eq:newvar}
  \rho(k,\eta) & =\frac{R(\kappa,\tau)}{4\pi g_{a\gamma}^{2}T_{QCD}^{2}k_{\mathrm{max}}^{2}},
  \quad
  h(k,\eta)=\frac{H(\kappa,\tau)}{2\pi g_{a\gamma}^{2}T_{QCD}^{2}k_{\mathrm{max}}^{3}},
  \nonumber
  \\
  \varphi_{k}(\eta) & =\frac{\sigma_{c}}{8\pi g_{a\gamma}k_{\mathrm{max}}^{4}}\Phi(\kappa,\tau),
  \quad
  \tau=\frac{2k_{\mathrm{max}}^{2}}{\sigma_{c}}\eta,
  \quad
  k=k_{\mathrm{max}}\kappa,
\end{align}
where $0\leq\tau\leq\tau_{\mathrm{max}}=2k_{\mathrm{max}}^{2}\tilde{M}_{\mathrm{Pl}}/T_{\mathrm{now}}$,
$\kappa_{m}<\kappa<1$ in the magnetic spectra, whereas $0<\kappa<1$
in the axion spectrum, $T_{\mathrm{now}}=2.7\,\text{K}$ is the current
universe temperature, and $\kappa_{m}=k_{\mathrm{min}}/k_{\mathrm{max}}$.
Using Eq.~(\ref{eq:newvar}), Eqs.~(\ref{eq:speceqmean}) and~(\ref{eq:axeqmean})
are rewritten in the form,
\begin{align}\label{eq:sysdmls}
  \frac{\partial R}{\partial\tau}= & -\kappa^{2}R+
  \kappa^{2}H \int_{0}^{1}\kappa'^{2}\mathrm{d}\kappa'
  \left[
    \frac{\partial\Phi(\kappa',\tau)}{\partial\tau}+\frac{\kappa'^{2}}{3}\Phi(\kappa',\tau)
  \right],
  \nonumber
  \\
  \frac{\partial H}{\partial\tau}= & -\kappa^{2}H+
  R\int_{0}^{1}\kappa'^{2}\mathrm{d}\kappa'
  \left[
    \frac{\partial\Phi(\kappa',\tau)}{\partial\tau}+\frac{\kappa'^{2}}{3}\Phi(\kappa',\tau)
  \right],
  \nonumber
  \\
  \frac{\partial^{2}\Phi}{\partial\tau^{2}}= & -\frac{2}{\tilde{a}}\frac{\mathrm{d}\tilde{a}}{\mathrm{d}\tau}\frac{\partial\Phi}{\partial\tau}-
  \left(
    \kappa^{2}\mathcal{K}^{2}+\tilde{a}^{2}\mu^{2}
  \right)
  \Phi
  \nonumber
  \\
  & +
  \frac{1}{\tilde{a}^{2}}
  \bigg\{
    \int_{\kappa_{m}}^{1}\kappa'^{2}H(\kappa',\tau)\mathrm{d}\kappa'
    \notag
    \\
    & -
    \int_{\kappa_{m}}^{1}R(\kappa',\tau)\mathrm{d}\kappa'\int_{0}^{1}\kappa'^{2}\mathrm{d}\kappa'
    \left[
      \frac{\partial\Phi(\kappa',\tau)}{\partial\tau}+\frac{\kappa'^{2}}{3}\Phi(\kappa',\tau)
    \right]
  \bigg\},
\end{align}
where $\mu=\tfrac{\sigma_{c}m}{2k_{\mathrm{max}}^{2}T_{\mathrm{QCD}}}$
is the effective axion mass, $\mathcal{K}=\sigma_{c}/2k_{\mathrm{max}}$,
$\tilde{a}=1+\lambda\tau$, and $\lambda=\tfrac{\sigma_{c}T_{\mathrm{QCD}}}{2k_{\mathrm{max}}^{2}\tilde{M}_{\mathrm{Pl}}}$.

The initial condition for the new variables in Eq.~(\ref{eq:newvar})
reads
\begin{equation}\label{eq:inicondRH}
  R(\kappa,0)=C\kappa^{n},
  \quad
  H(\kappa,0)=q\frac{R(\kappa,0)}{\kappa},
  \quad
  C=\frac{2\pi g_{a\gamma}^{2}T_{\mathrm{QCD}}^{2}k_{\mathrm{max}}(n+1)B_{c}^{(0)2}}{(1-\kappa_{m}^{n+1})},
\end{equation}
where $B_{c}^{(0)}=B_{\mathrm{now}}/T_{\mathrm{now}}^{2}=3.6\times10^{-4}$
is the seed conformal magnetic field which corresponds to the present
day magnetic field $B_{\mathrm{now}}=10^{-9}\,\text{G}$. Such a strength
is within the current constrains on the extragalactic magnetic field
obtained in Ref.~\cite{Pal22} basing on the cosmic rays observations.
However, lower bounds on cosmic magnetic fields are considered in
Ref.~\cite{NerSemKal21}. The initial condition for the dimensionless
axion wavefunction is
\begin{align}\label{eq:inicondPhi}
  \Phi(\kappa,0) & =\frac{32\pi^{3/2}\alpha_{\mathrm{em}}k_{\mathrm{max}}^{4}}{\sigma_{c}\delta^{3}}\exp
  \left(
    -\kappa^{2}\frac{k_{\mathrm{max}}^{2}}{\delta^{2}}
  \right),
  \nonumber
  \\
  \frac{\partial\Phi(\kappa,0)}{\partial\tau} & =\sqrt{\mathcal{K}^{2}\kappa^{2}+\mu^{2}}\Phi(\kappa,0),
\end{align}
where we take into account that~\cite{Cha21} $g_{a\gamma}\approx\tfrac{\alpha_{\mathrm{em}}}{2\pi f_{a}}$
and $\alpha_{\mathrm{em}}=7.3\times10^{-3}$ is the fine structure
constant.

\section{Results\label{sec:RES}}

In this section, we show the results of the numerical solution of
the system in Eq.~(\ref{eq:sysdmls}) with the initial condition
in Eqs.~(\ref{eq:inicondRH}) and~(\ref{eq:inicondPhi}).

\begin{figure}
  \centering
  \subfigure[]
  {\label{fig:inhomhoma}
  \includegraphics[scale=.33]{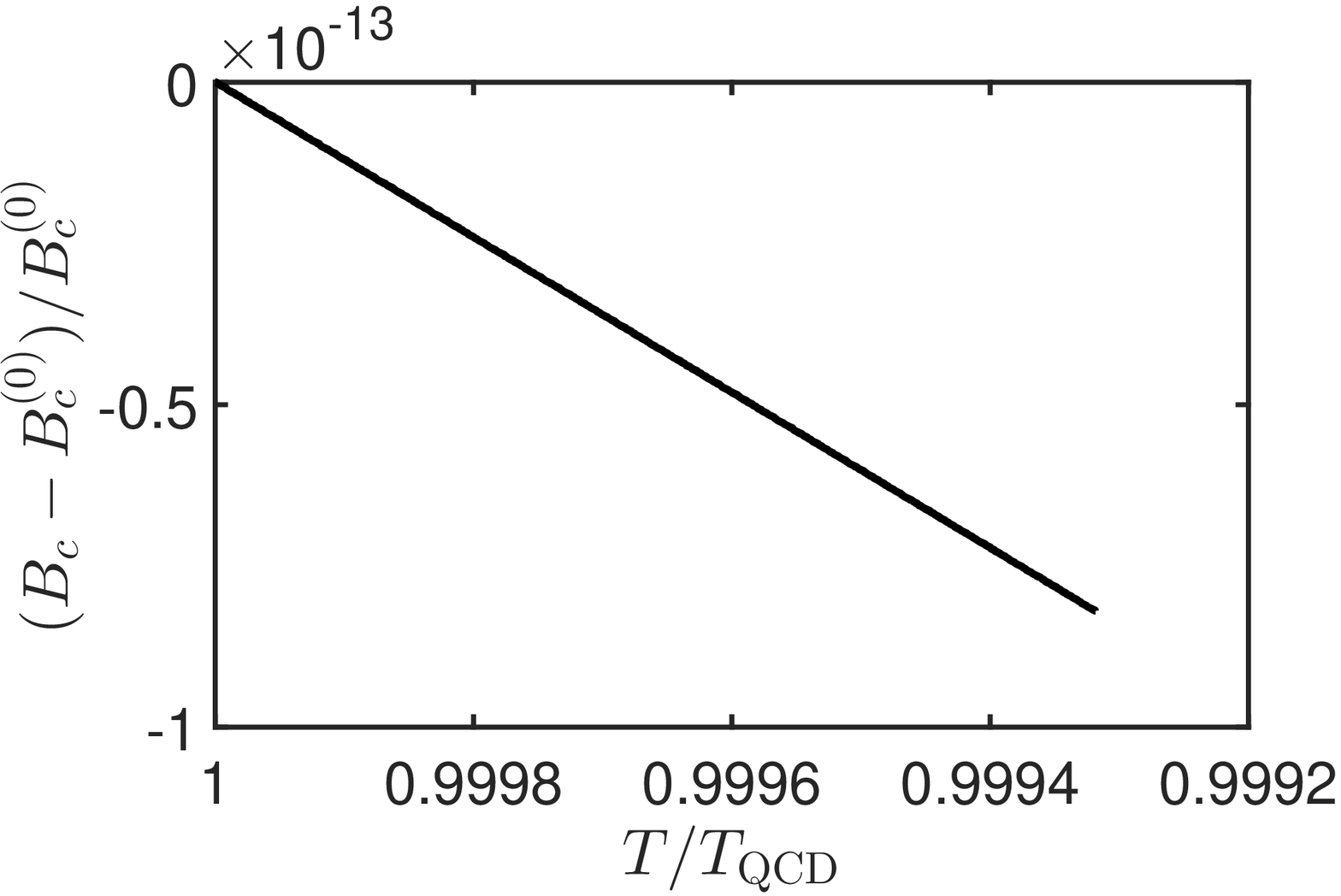}}
  \hskip-.4cm
  \subfigure[]
  {\label{fig:inhomhomb}
  \includegraphics[scale=.33]{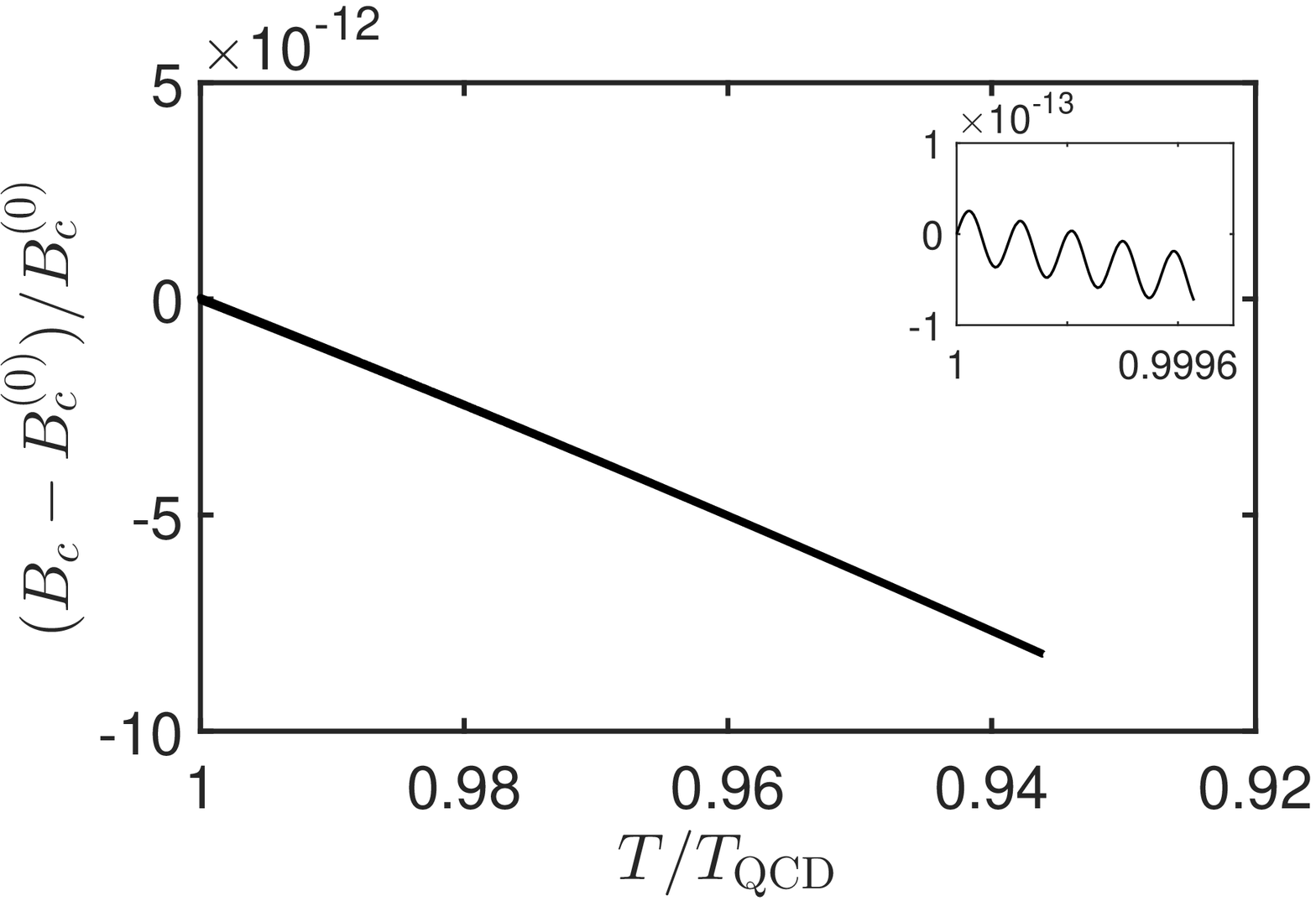}}
  \\
  \subfigure[]
  {\label{fig:inhomhomc}
  \includegraphics[scale=.33]{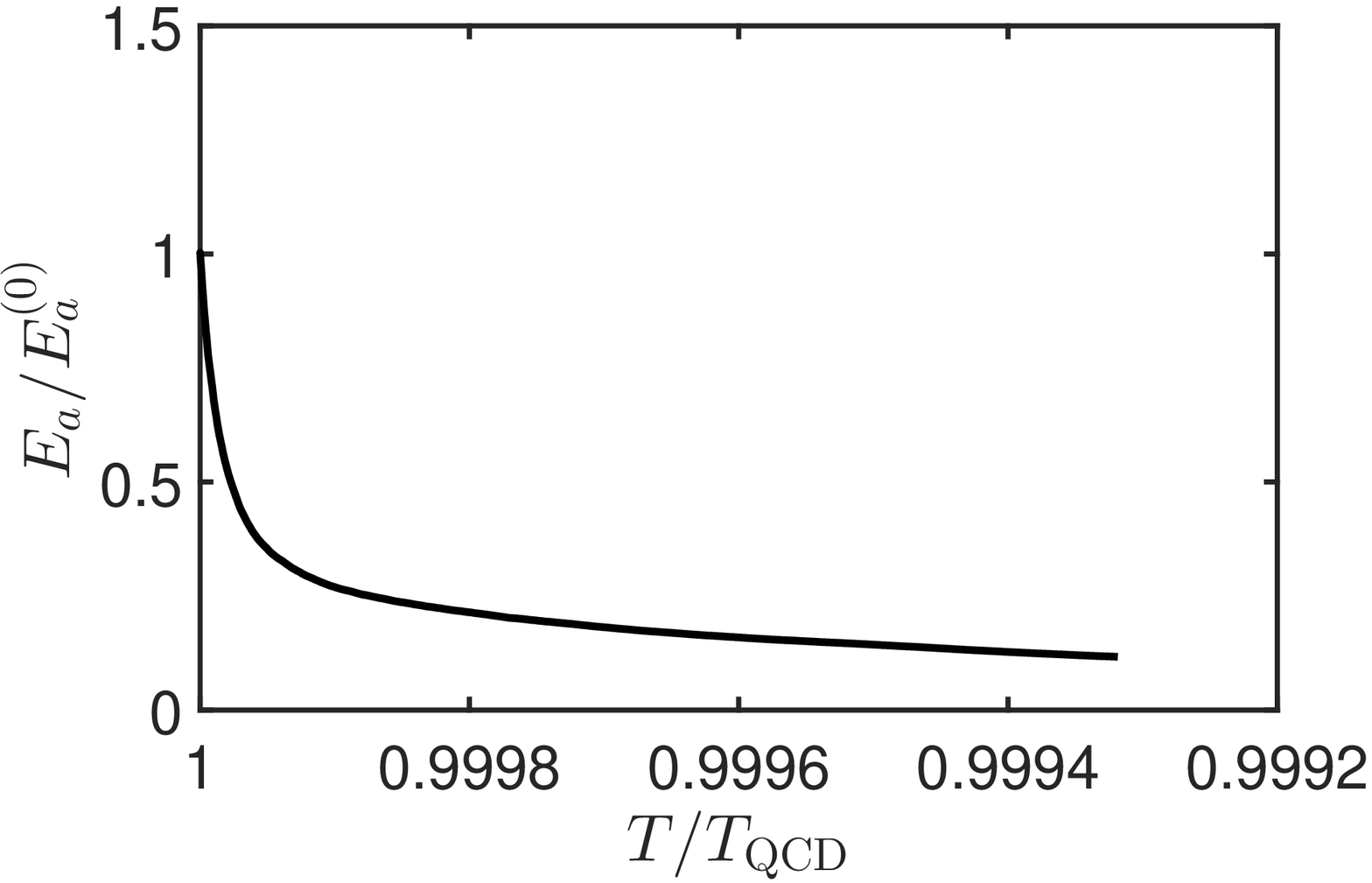}}
  \hskip-.4cm
  \subfigure[]
  {\label{fig:inhomhomd}
  \includegraphics[scale=.33]{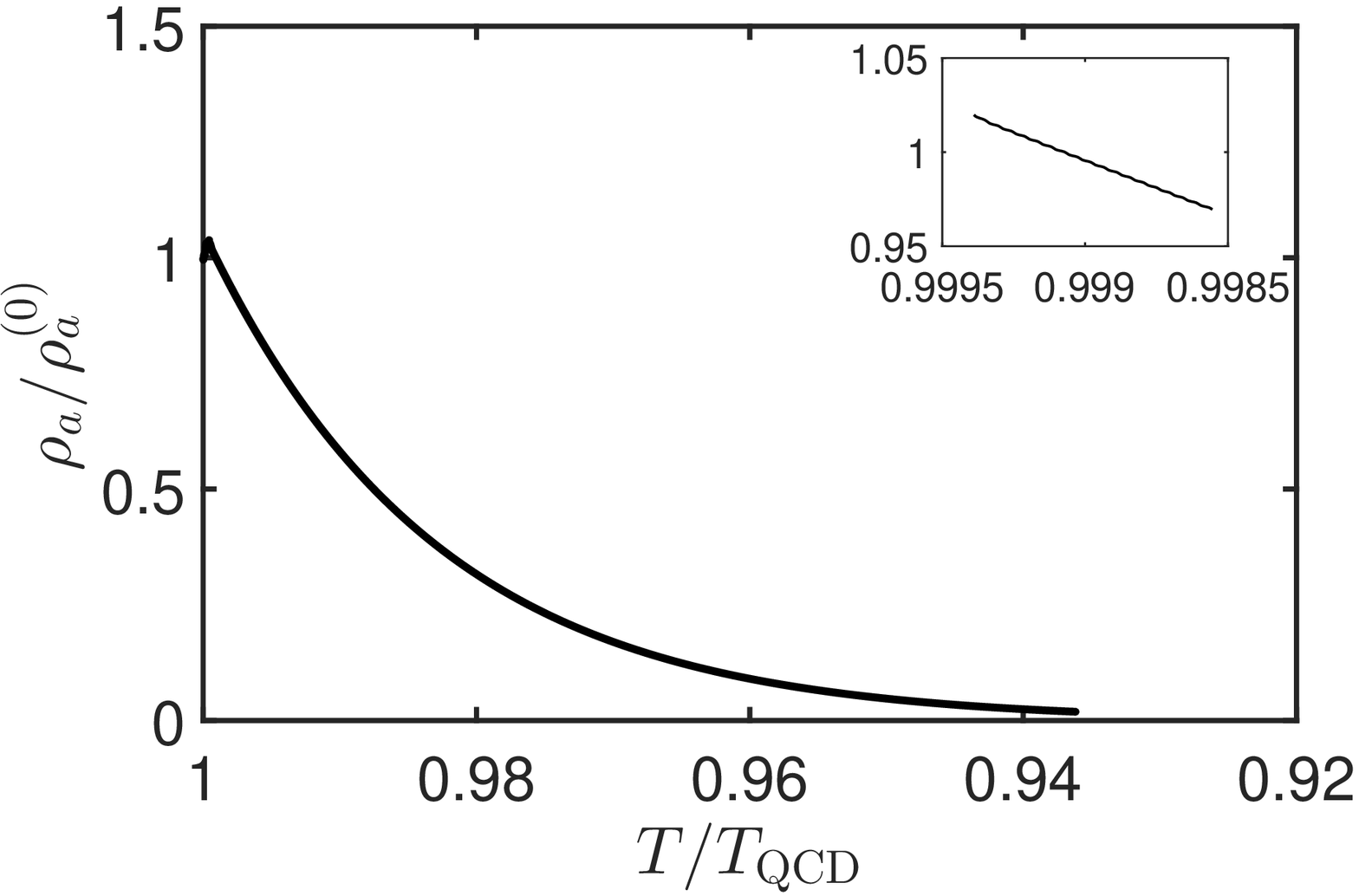}}
  \protect
\caption{The observable parameters of the system while it evolves in the universe cooling down from $T_{\mathrm{QCD}}$.
We take that $k_{\mathrm{max}}=10^{-10}$, $B_{c}^{(0)}=3.6\times10^{-4}$
($B_{\mathrm{now}}=10^{-9}\,\text{G}$), $m=10^{-6}\,\text{eV}$,
$f_{a}=10^{12}\,\text{GeV}$, and $q=1$. We choose the seed Kolmogorov spectrum. Panels~(a) and~(c): the case of the inhomogeneous axion with the initial condition in Eq.~\eqref{eq:inicondPhi} and $\delta=10^{-10}$. Panels~(b) and~(d): the situation of the homogeneous axion corresponding to $\delta=0$ in Eq.~\eqref{eq:inicondPhi} and $\varphi_0=f_a$. The insets in panels~(b) and~(d) show the corresponding quantities at small evolution times. Panels~(a)
and~(b): the evolution of the magnetic field. Panel~(c):
the behavior of the total axion energy in Eq.~(\ref{eq:toten}) of the inhomogeneous axion. Panel~(d): the evolution of the energy density $\rho_a$ of the homogeneous axion.\label{fig:inhomhom}}
\end{figure}

The system analogous to that in Eq.~(\ref{eq:sysdmls}) was studied
in Ref.~\cite{DvoSem20} for a spatially homogeneous axion. Here,
we are mainly focused on the influence of a nontrivial seed spectrum
of an axion on the evolution of the system. That is why we compare the inhomogeneous axion MHD with the homogeneous one. As we mentioned in Sec.~\ref{sec:INICOND}, the parameter $\delta \sim k_\mathrm{max}$. In Fig.~\ref{fig:inhomhom}, we compare the cases of $\delta = 10^{-10}$ (the inhomogeneous axion) and $\delta = 0$ (the homogeneous axion). The rest of the parameters and the initial condition are the same for both the inhomogeneous and homogeneous cases.

Unfortunately, we can proceed in numerical simulation only for $T\lesssim T_{\mathrm{QCD}}$ for inhomogeneous axions because of the very rapid oscillations of the axion wavefunction (see Fig.~\ref{fig:inhomphi} below). We checked the dependence of the evolution of the system on the seed magnetic helicity $q$ defined in Sec.~\ref{sec:INICOND}. It turns out to be negligible. Thus, we set $q=1$ which corresponds to the maximally helical fields. 

The measurable quantities in the axion MHD are the conformal magnetic field,
\begin{equation}\label{eq:MF}
  B_c = 
  2
  \left[
    \int_{k_\mathrm{min}}^{k_\mathrm{max}}
    \mathrm{d}k \rho(k,\eta)
  \right]^{1/2},
\end{equation}
and the total axion energy in the cooling
universe, $E_{a}\propto\smallint\mathrm{d}^{3}xT_{00}$,
\begin{equation}\label{eq:toten}
  E_{a}=\int_{0}^{1}\kappa^{2}\mathrm{d}\kappa
  \left[
    \left(
      \frac{\partial\Phi}{\partial\tau}
    \right)^{2}+
    \left(
      \kappa^{2}\mathcal{K}^{2}+\tilde{a}^{2}\mu^{2}
    \right)\Phi^{2}
  \right],
\end{equation}
where $T_{00}$ is the zeroth component of the axion energy-momentum
tensor. We normalize $E_{a}$ on its initial value $E_{a}^{(0)}$. In case of a homogeneous axion, we deal with the axion energy density $\rho_a = (\partial_\tau \Phi)^2 + \tilde{a}^2 \mu^2 \Phi^2$, which is also normalized on $\rho_a^{(0)} = (\partial_\tau \Phi_0)^2 + \mu^2 \Phi_0^2$.

We show $B_c$ versus $T$ in the universe cooling down from $T_\mathrm{QCD}$ in Figs.~\ref{fig:inhomhoma} and~\ref{fig:inhomhomb} in inhomogeneous and homogeneous cases respectively. The magnetic field decays in both cases, mainly because of the diffusion terms, $\propto -\kappa^2 R$ and $\propto -\kappa^2 H$, in Eq.~\eqref{eq:sysdmls}.

The evolution time in Fig.~\ref{fig:inhomhoma} approximately corresponds to that in the inset in Fig.~\ref{fig:inhomhomb}. We can see the ripple in the inset in Fig.~\ref{fig:inhomhomb} (homogeneous axion), whereas the line in Fig.~\ref{fig:inhomhoma} (inhomogeneous axion) is smooth. This ripple is not noticeable on great evolution times in Fig.~\ref{fig:inhomhomb}. Comparing Fig.~\ref{fig:inhomhoma} and the inset in Fig.~\ref{fig:inhomhomb}, we can see that the inhomogeneous case approximately corresponds to the mean value of the homogeneous axion wavefunction.

We depict $E_a$ in Fig.~\ref{fig:inhomhomc} and $\rho_a$ in Fig.~\ref{fig:inhomhomd} versus $T$. By the end of the evolution, both quantities are about $0.1$ from their initial values. However, the decay in the inhomogeneous case in Fig.~\ref{fig:inhomhomc} happens much faster than in the homogeneous situation in Fig.~\ref{fig:inhomhomd}. Such a behavior is owing to the additional term $\propto \kappa'^2 \Phi$ in the integrand in the last line of Eq.~\eqref{eq:sysdmls}. Therefore the damping of the inhomogeneous axion is greater effectively.

In Fig.~\ref{fig:inhomphi}, we show the evolution of the axion wavefunction integrated over the spectrum,
\begin{equation}\label{eq:wfav}
  \bar{\varphi}(\eta)=\int_{0}^{k_{\mathrm{max}}}\mathrm{d}k\,k^{2}\varphi_{k}(\eta),
\end{equation}
which is again normalized on its initial value at $T_{\mathrm{QCD}}$. One can see that it oscillates vary rapidly. As mentioned above, this fact makes the solution of the system in Eq.~\eqref{eq:sysdmls} more complicated than for the homogeneous axion studied in Ref.~\cite{DvoSem20}. Moreover, as one can see in the inset in Fig.~\ref{fig:inhomphi}, there are both rapid oscillations and a sort of beatings in the evolution of $\bar{\varphi}$. These beatings appear since we integrate different wavenumbers over the spectrum in Eq.~\eqref{eq:wfav}.

\begin{figure}
  \centering
  \includegraphics[scale=.33]{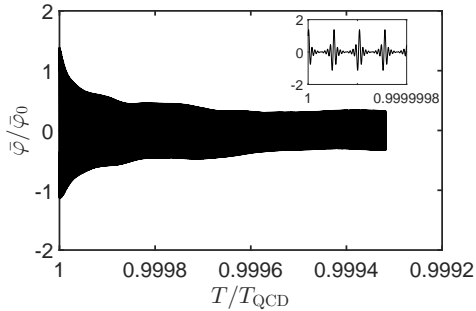}
  \protect
\caption{The evolution of the wavefunction of the inhomogeneous axion integrated over the spectrum
in Eq.~(\ref{eq:wfav}). The parameters of the system are the same as in Figs.~\ref{fig:inhomhoma} and~\ref{fig:inhomhomc}.\label{fig:inhomphi}}
\end{figure}

We take that $k_{\mathrm{max}}=10^{-10}$ in our simulations. It means
that the present length scale obeys $l_{\mathrm{now}}>(k_{\mathrm{max}}T_{\mathrm{now}})^{-1}\approx10^{-2}R_{\odot}$,
where $R_{\odot}$ is the solar radius. Thus the perturbations of
the axion field considered in our work, can result in the formation
of hypothetical astronomical objects like Bose stars; cf. Ref.~\cite{KolTka93}.

\subsection{Magnetic field instability\label{subsec:TOY}}

We showed above that there are certain differences in the behavior of the measurable quantities in MHD in the presence of inhomogeneous and homogeneous axions. Nevertheless, these differences are smeared at low temperatures $\sim T_\mathrm{now}$. The axion wavefunction contributes to the $\alpha$-dynamo parameter in Eq.~\eqref{eq:alpha0}, which is known to result in the instability of the magnetic field. However, we did not observe this instability in the cases studied above. The behavior of $B_c$ was mainly affected by the magnetic diffusion.

It happens since we take that the initial axion wavefunction is $\varphi_0 \sim f_a$ at $T_\mathrm{QCD}$. This initial condition is realistic (see, e.g., Ref.~\cite{LonVac15}). However, the impact of the axion on the magnetic field is small in this case.

Now we consider the situation when the initial axion wavefunction is increased. Namely, we take that $\varphi_{k}^{(0)}=8\pi^{3/2} \Xi f_{a}\exp(-k^{2}/\delta^{2})/\delta^{3}$, where the factor $\Xi = 5\times 10^7$. We also consider the homogeneous axion with $\varphi_0 = \Xi f_{a}$, where $\Xi$ has the same value as in the inhomogeneous case. The initial axion wavefunctions, which are greater than $f_a$, were mentioned in Ref.~\cite{LonVac15} to potentially influence magnetic fields. We shall see soon that it is the case. The rest of the parameters and the initial condition are the same as in Fig.~\ref{fig:inhomhom}.

\begin{figure}
  \centering
  \subfigure[]
  {\label{fig:toyinhomhoma}
  \includegraphics[scale=.33]{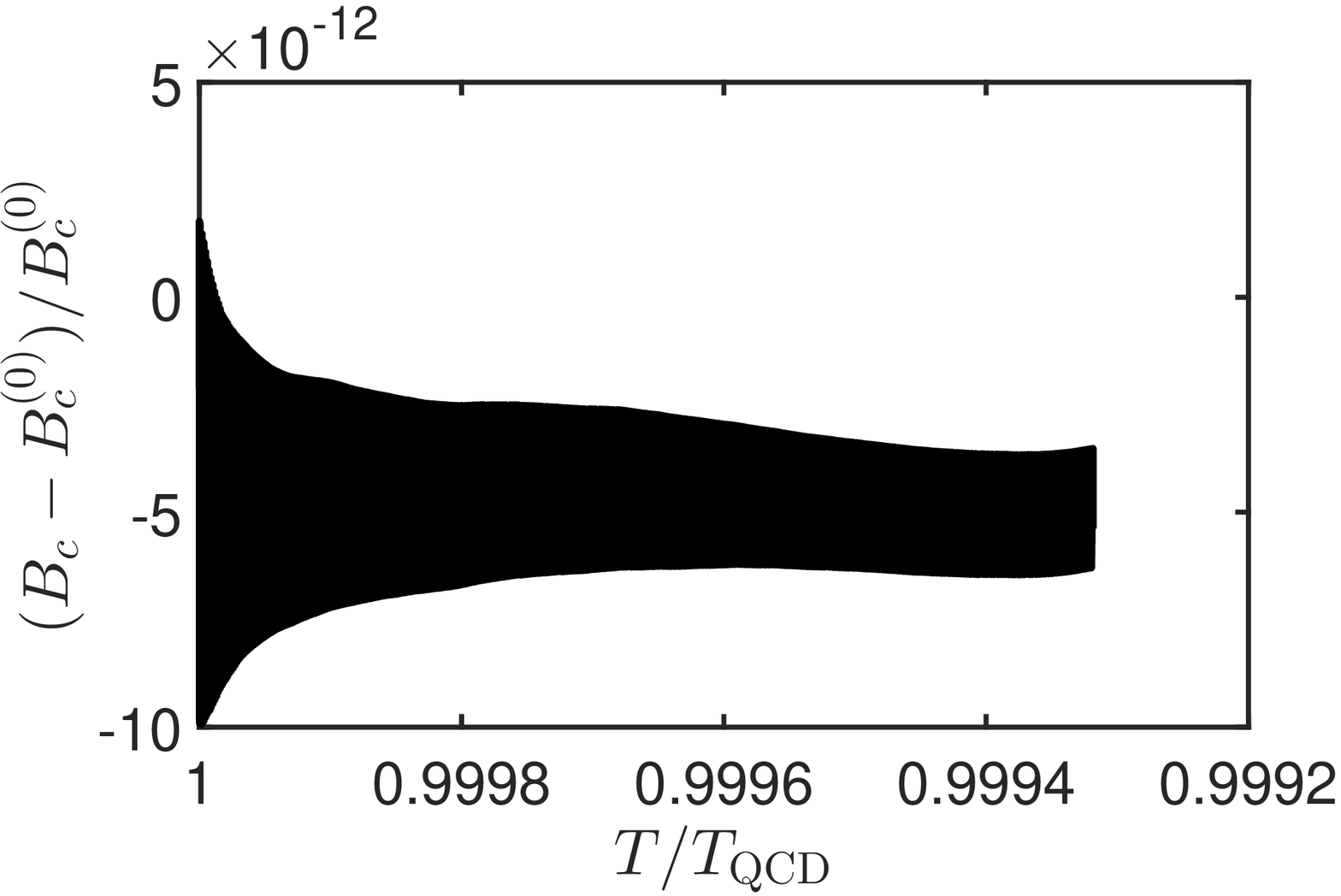}}
  \hskip-.4cm
  \subfigure[]
  {\label{fig:toyinhomhomb}
  \includegraphics[scale=.33]{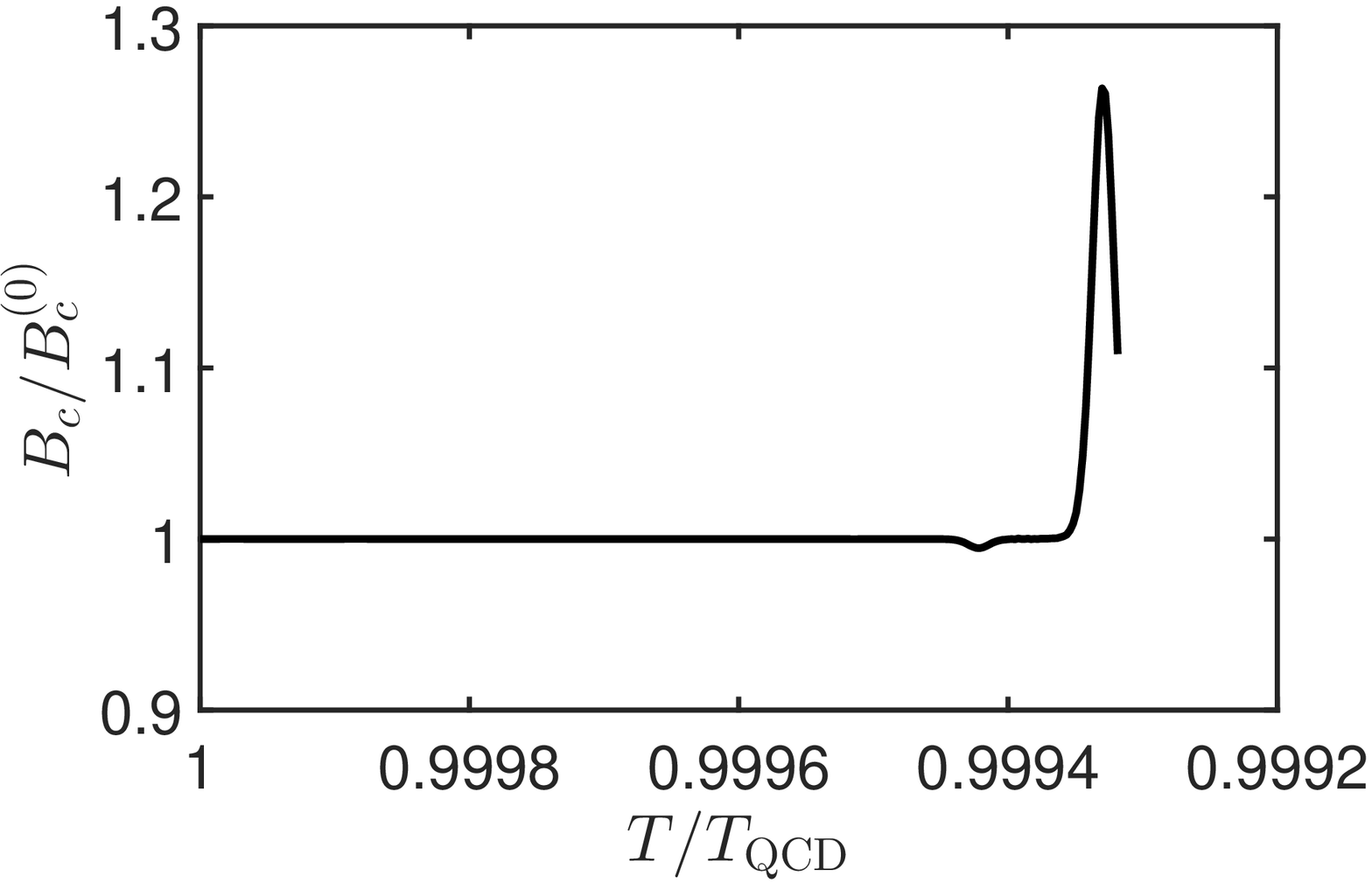}}
  \protect
\caption{The evolution of the magnetic field for (a) the inhomogeneous axion; and (b) the homogeneous one. The parameters are the same as in Fig.~\ref{fig:inhomhom} except for the initial axion wavefunction which is amplified by the factor $\Xi = 5\times10^7$ in both panels
 now.\label{fig:toyinhomhom}}
\end{figure}

In Fig.~\ref{fig:toyinhomhom}, we show the evolution of the magnetic field defined in Eq.~\eqref{eq:MF}. We study the inhomogeneous and homogeneous cases in the same time interval. One can see in Fig.~\ref{fig:toyinhomhomb} that the magnetic field becomes unstable. The evolution of the magnetic field in the inhomogeneous case shown in Fig.~\ref{fig:toyinhomhoma} does not reveal the instability.

\begin{figure}
  \centering
  \subfigure[]
  {\label{fig:1-10a}
  \includegraphics[scale=.33]{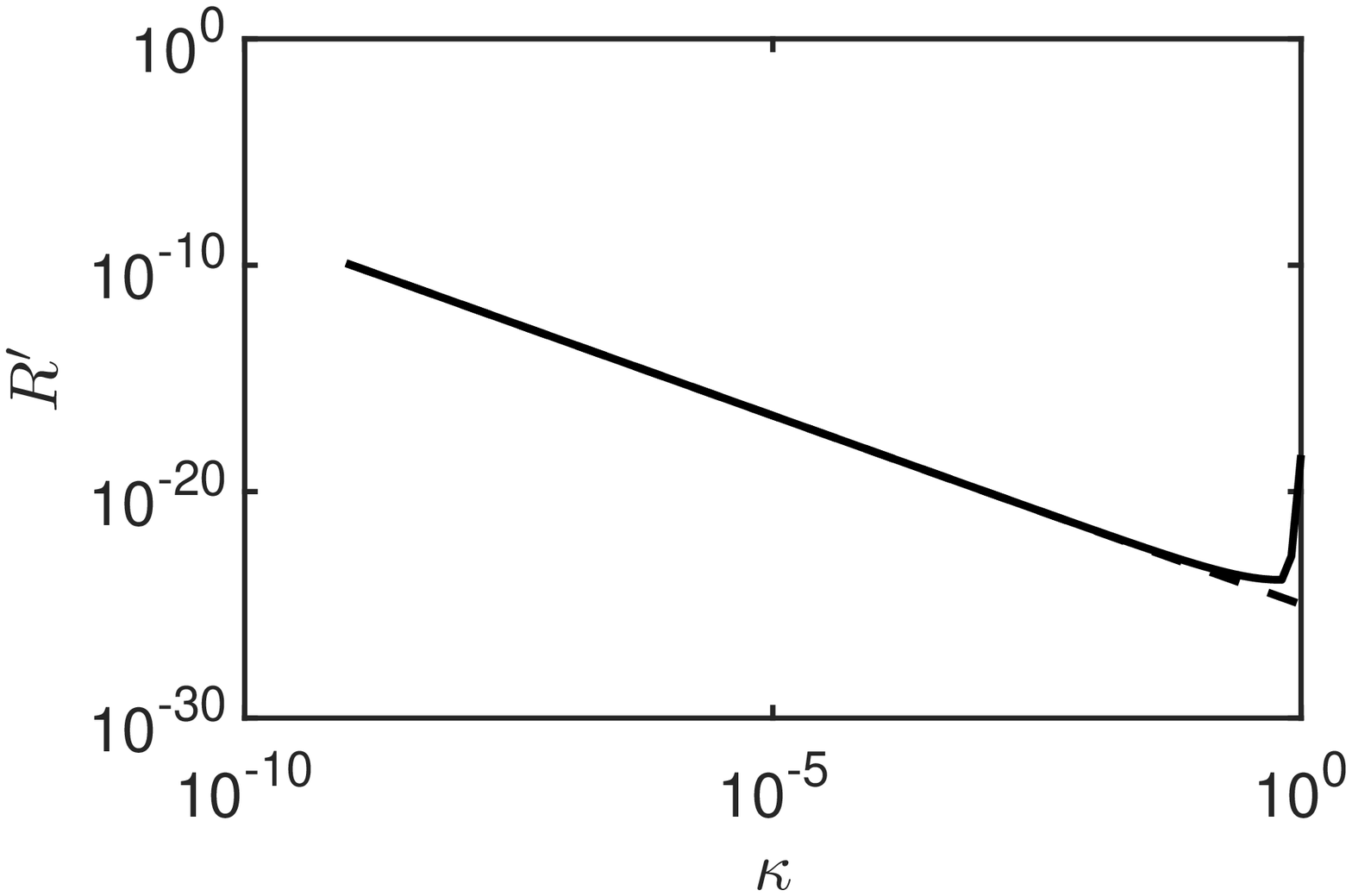}}
  \hskip-.5cm
  \subfigure[]
  {\label{fig:1-10b}
  \includegraphics[scale=.33]{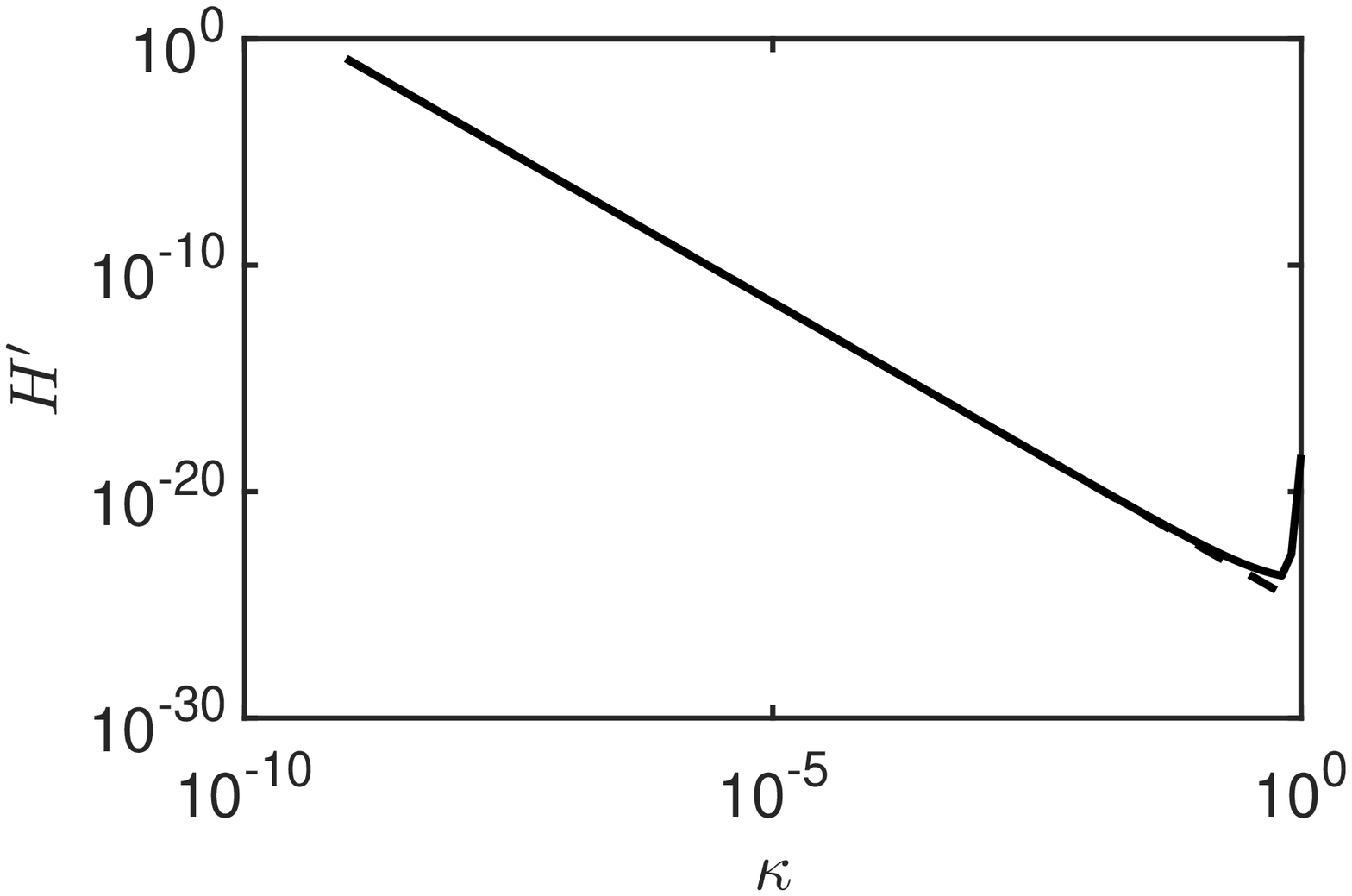}}
  \protect
\caption{The spectra of the densities of (a) the magnetic energy; and (b) the magnetic helicity in case of the homogeneous axion. The parameters of the system and the initial condition are the same as in Fig.~\ref{fig:toyinhomhomb}. Solid lines are final spectra at the end of the system evolution. Dashed lines are the seed Kolmogorov spectra.\label{fig:homspec}}
\end{figure}

To illustrate the development of the instability of the magnetic field in the homogeneous case, we plot the spectra of the densities of the magnetic energy and the magnetic helicity in Fig.~\ref{fig:homspec}. Unlike Eq.~\eqref{eq:newvar}, here we use different new variables, $R'(\kappa,\tau) = g_{a\gamma}^2 T_\mathrm{QCD}^2 \rho(k,\eta)/k_\mathrm{max}$ and $H'(\kappa,\tau) = g_{a\gamma}^2 T_\mathrm{QCD}^2 h(k,\eta)/2$. We studied the evolution of magnetic fields with $\Xi < 5\times 10^7$ and did not reveal the magnetic field instability, at least in the considered time interval.

In Fig.~\ref{fig:homspec}, by dashed lines, we show the seed Kolmogorov spectra which correspond to $T_\mathrm{QCD}$. We also depict the final spectra which are at the end of the evolution of the system. One can see that parts of the spectra at great momenta, i.e. at small scales, increase. One can say that it is equivalent to the direct energy cascade triggered by the homogeneous axion.

The $\alpha$-dynamo parameter in Eq.~\eqref{eq:alpha0} has two axion contributions in the integrand: (i) $\propto \varphi'_k$; and (ii) $\propto k^2 \varphi_k$. Despite we take a greater axion wavefunction, the inhomogeneous  axion oscillates more rapidly; cf. Fig.~\ref{fig:inhomphi}. Thus, the term (i) prevails over the term (ii). Since the term (i) is sign changing, the consideration of the inhomogeneous axion prevents the development of the magnetic field instability.

\section{Conclusion\label{sec:CONCL}}

In the present work, we have studied the evolution of the system which
consists of axions and magnetic fields. Both axions and magnetic
fields are spatially inhomogeneous. The evolution of such a system
was considered in the early universe cooling down from the QCD phase
transition, with the universe expansion rate being accounted for exactly.

First, in Sec.~\ref{sec:MAGNAX}, we have started with the general
equations for an electromagnetic field, which interacts with axions,
in an arbitrary curved spacetime. Then, taking the FRW metric, we have
derived the equations for the conformal electric and magnetic fields.
Basing on this result, we have obtained the general induction Eq.~(\ref{eq:Indgen})
for the magnetic field, which accounts for the spatial distribution
of axions. This equation has been further modified to Eq.~(\ref{eq:Ind})
in the isotropic medium. Then, we have obtained the general Eq.~(\ref{eq:speceqgen})
for the spectra of the magnetic energy and the magnetic helicity.

To describe the evolution of axions in the presence of an electromagnetic
field, in Sec.~\ref{sec:AXMAGN}, we have considered the general
Eq.~(\ref{eq:axcurved}) for $\varphi$ in a curved spacetime. This
equation has been rewritten in the Fourier space in Eq.~(\ref{eq:axfour}).
An arbitrary magnetic field contributes to the evolution of axions
via the conformal time derivative of the magnetic helicity spectrum
in Eq.~(\ref{eq:helfkp}).

In order to get the closed system of the evolution equations for a
magnetic field and an axion, in Sec.~\ref{sec:MEAN}, we have applied
the mean field approximation which consists in the consideration of zero mode perturbations. In this situation, we could define the
mean spectra, $\rho(k,\eta)$ and $h(k,\eta)$, in Eq.~(\ref{eq:rho0h0}),
as well as derive Eqs.~(\ref{eq:speceqmean}) and~(\ref{eq:axeqmean}).
Formally, Eqs.~(\ref{eq:speceqmean}) and~(\ref{eq:axeqmean}) coincide
with those obtained in Ref.~\cite{DvoSem20}. However, they account
for the nontrivial spectrum of the $\alpha$-dynamo parameter.

In Sec.~\ref{sec:INICOND}, we have chosen the initial conditions
and have defined all the parameters in Eqs.~(\ref{eq:speceqmean})
and~(\ref{eq:axeqmean}). We have considered the evolution of the
system below the QCD phase transition when the axion mass takes its
vacuum value. It should be noted that there is not so much information
about the seed spectrum of axions. We have adopted the initial Gaussian
distribution for $\varphi_{k}$ which is characterized by the phenomenological
parameter $\delta$. One can reproduce, e.g., the zero mode case,
discussed in Refs.~\cite{LonVac15,DvoSem20}, by setting $\delta\to0$.
For the magnetic field, we have chosen the seed Kolmogorov spectrum
with realistic parameters. Finally, we have rewritten the evolution
equations in the form convenient for numerical simulations; cf. Eq.~(\ref{eq:sysdmls}).

The numerical solution of Eq.~(\ref{eq:sysdmls}) has been present
in Sec.~\ref{sec:RES}. We have been mainly interested in the dependence
of the results on the seed axion spectrum. For this purpose we have compared the cases of the inhomogeneous, with $\delta = 10^{-10}$, and the homogeneous, with $\delta = 0$, axions. In both cases, the magnetic field evolution is different only in small time intervals. It is smooth for inhomogeneous axions and reveal a ripple for homogeneous axions; cf. Figs.~\ref{fig:inhomhoma} and~\ref{fig:inhomhomb}. Globally, the evolution of magnetic fields is driven mainly by the diffusion terms in Eq.~\eqref{eq:sysdmls}. The energy of the inhomogeneous axion decays faster than for the homogeneous one; cf. Figs.~\ref{fig:inhomhomc} and~\ref{fig:inhomhomd}.

The possibility to get the magnetic field instability has been considered in Sec.~\ref{subsec:TOY}. For this purpose we have taken the initial axion wavefunction amplified by the factor $\Xi = 5\times 10^7$. The impact of axions on the magnetic field was mentioned in Ref.~\cite{LonVac15} to be possible if the initial axion wavefunction is greater than $f_a$. We have obtained that the magnetic field can be unstable in the case of a homogeneous axion; cf. Fig.~\ref{fig:toyinhomhomb}. This instability is washed out for the inhomogeneous axion (see Fig.~\ref{fig:toyinhomhoma}) owing to the very rapid sign alteration of the $\alpha$-dynamo parameter.

\section*{Acknowledgments}

I am thankful to P.M.~Akhmetev and S.V.~Troitsky for useful
discussions.


\begin{thebibliography}{50}

\bibitem{PecQui77}
  R.D.~Peccei, H.R.~Quinn,
  CP Conservation in the Presence of Pseudoparticles,
  Phys. Rev. Lett. 38 (1977) 1440--443.

\bibitem{Abe20}
  C.~Abel, et al.,
  Measurement of the permanent electric dipole moment of the neutron,
  Phys. Rev. Lett. 124 (2020) 081803,
  arXiv:2001.11966.

\bibitem{Luz20}
  L.~Di Luzio, M.~Giannotti, E.~Nardi, L.~Visinelli,
  The landscape of QCD axion models,
  Phys. Rept. 870 (2020) 1--117,
  arXiv:2003.01100.

\bibitem{Gra15}
  P.W.~Graham, I.G.~Irastorza, S.K.~Lamoreaux, A.~Lindner, K.A.~van Bibber,
  Experimental searches for the axion and axion-like particles,
  Annu. Rev. Nucl. Part. Sci. 65 (2015) 485--514,
  arXiv:1602.00039.

\bibitem{Apr20}
  XENON Collaboration,
  Excess electronic recoil events in XENON1T,
  Phys. Rev. D 102 (2020) 072004,
  arXiv:2006.09721.

\bibitem{Ana17}
  CAST Collaboration,
  New CAST limit on the axion-photon interaction,
  Nature Phys. 13 (2017) 584--590,
  arXiv:1705.02290.

\bibitem{Bus21}
  M.~Buschmann, R.T.~Co, C.~Dessert, B.R.~Safdi,
  Axion Emission Can Explain a New Hard X-Ray Excess from Nearby Isolated Neutron Stars,
  Phys. Rev. Lett. 126 (2021) 021102,
  arXiv:1910.04164.

\bibitem{SemYou21}
  Y.K.~Semertzidis, S.~Youn,
  Axion Dark Matter: How to see it?,
  Sci. Adv. 8 (2022) eabm9928,
  https://doi.org/10.1126/sciadv.abm9928,
  arXiv:2104.14831.

\bibitem{Cha21}
  F.~Chadha-Day, J.~Ellis, D.J.E.~Marsh,
  Axion Dark Matter: What is it and Why Now?,
  Sci. Adv. 8 (2022) eabj3618,
  https://doi.org/10.1126/sciadv.abj3618,
  arXiv:2105.01406.

\bibitem{DinFis83}
  M.~Dine, W.~Fischler,
  The Not So Harmless Axion,
  Phys. Lett. B 120 (1983) 137--141.

\bibitem{Mar16}
  D.J.E.~Marsh,
  Axion Cosmology,
  Phys. Rept. 643 (2016) 1--79,
  arXiv:1510.07633.

\bibitem{KolTka93}
  E.W.~Kolb, I.I.~Tkachev,
  Axion miniclusters and Bose stars,
  Phys. Rev. Lett. 71 (1993) 3051--3054,
  hep-ph/9303313.

\bibitem{KhlSakSok99}
  M.Yu.~Khlopov, A.S.~Sakharov, D.D.~Sokoloff,
  The nonlinear modulation of the density distribution in standard
  axionic CDM and its cosmological impact,
  Nucl. Phys. B (Proc. Suppl.) 72 (1999) 105--109.

\bibitem{EnaParSch17}
  J.~Enander, A.~Pargner, T.~Schwetz,
  Axion minicluster power spectrum and mass function,
  J. Cosmol. Astropart. Phys. 12 (2017) 038,
  arXiv:1708.04466.

\bibitem{LonVac15}
  A.~Long, T.~Vachaspati,
  Implications of a primordial magnetic field for magnetic monopoles, axions, and Dirac neutrinos,
  Phys. Rev. D 91 (2015) 103522,
  arXiv:1504.03319.

\bibitem{DvoSem20}
  M.~Dvornikov, V.B.~Semikoz,
  Evolution of axions in the presence of primordial magnetic fields,
  Phys. Rev. D 102 (2020) 123526,
  arXiv:2011.12712.

\bibitem{FukKhaWar08}
  K.~Fukushima, D.E.~Kharzeev, H.J.~Warringa,
  The chiral magnetic effect,
  Phys. Rev. D 78 (2008) 074033,
  arXiv:0808.3382.

\bibitem{Bra17}
  A.~Brandenburg, J.~Schober, I.~Rogachevskii, T.~Kahniashvili, A.~Boyarsky, J.~Fr\"ohlich, O.~Ruchayskiy, N.~Kleeorin,
  The Turbulent Chiral Magnetic Cascade in the Early Universe,
  Astrophys. J. Lett. 845 (2017) L21,
  arXiv:1707.03385.

\bibitem{BraEnqOle96}
  A.~Brandenburg, K.~Enqvist, P.~Olesen,
  Large-scale magnetic fields from hydromagnetic turbulence in the very early universe,
  Phys. Rev. D 54 (1996) 1291--1300,
  astro-ph/9602031.

\bibitem{BoyFroRuc12}
  A.~Boyarsky, J.~Fr\"ohlich, O.~Ruchayskiy,
  Self-Consistent Evolution of Magnetic Fields and Chiral Asymmetry in the Early Universe,
  Phys. Rev. Lett. 108 (2012) 031301,
  arXiv:1109.3350.

\bibitem{DvoSem14}
  M.~Dvornikov, V.B.~Semikoz,
  Instability of magnetic fields in electroweak plasma driven by neutrino asymmetries,
  J. Cosmol. Astropart. Phys. 05 (2014) 002,
  arXiv:1311.5267.

\bibitem{Oue19}
  J.L.~Ouellet, et al.,
  First Results from ABRACADABRA-10\,cm: A Search for Sub-$\mu$eV Axion Dark Matter,
  Phys. Rev. Lett. 122 (2019) 121802,
  arXiv:1810.12257.

\bibitem{Gra21}
  A.V.~Gramolin, D.~Aybas, D.~Johnson, J.~Adam, A.O.~Sushkov,
  Search for axion-like dark matter with ferromagnets,
  Nature Phys. 17 (2021) 79--84,
  arXiv:2003.03348.

\bibitem{Hus16}
  L.~Husdal,
  On Effective Degrees of Freedom in the Early Universe,
  Galaxies 4 (2016) 78,
  arXiv:1609.04979.

\bibitem{Pal22}
  A.~van Vliet, A.~Palladino, A.~Taylor, W.~Winter,
  Extragalactic magnetic field constraints from ultra-high-energy cosmic rays from local galaxies,
  Mon. Not. R. Astron. Soc. 510 (2022) 1289--1297,
  arXiv:2104.05732.

\bibitem{NerSemKal21}
  A.~Neronov, D.~Semikoz, O.~Kalashev,
  Limit on intergalactic magnetic field from ultra-high-energy cosmic ray hotspot in Perseus-Pisces region,
  arXiv:2112.08202.
  
\end{thebibliography}
\end{document}